# Use Energy Storage for Primary Frequency Control in Power Grids

Shutang You

***Abstract*— Frequency stability of power systems becomes more vulnerable with the increase of solar photovoltaic (PV). Energy storage provides an option to mitigate the impact of high PV penetration. Using the U.S. Eastern Interconnection (EI) and Texas Interconnection (ERCOT) power grid models, this paper investigates the capabilities of using energy storage to improve frequency response under high PV penetration. The study result helps to identify the potential and impact factors in utilizing energy storage to improve frequency response in high renewable penetration power grids.

## I. Introduction

Photovoltaic (PV) generation and wind power generation are increasing in power systems of many nations [1-5]. The retirement of conventional units and the increase of PV generation will deteriorate the frequency response capability of power grids. As PV inverters are typically operated at the Maximum Power Point, they usually can not generate extra power when the system frequency declines. To reserve PV headroom for frequency response, a trade-off should be made to balance the reliability benefit and the opportunity cost. Besides PV output reserve, energy storage (ES) is another option to improve the grid frequency response [6, 7].

With the decreasing price of energy storage systems, interconnection-level frequency control using power-electronics-interfaced energy storage has become economically feasible. Some literature has explored different control strategies for energy storage frequency control. For example, Ref. [8] applied droop control in energy storage and developed a new approach for optimizing the size and operation of energy storage to achieve the maximum profitability. Some studies found the response time of droop control is relatively long and it may result in large frequency deviations after major contingencies. To mitigate this drawback, Ref. [9] utilized the time derivative of frequency (ROCOF) to activate additional frequency control. Further, Ref. [10] combined inertia emulation with the conventional droop control in energy storage frequency regulation. To coordinate the charging of distributed energy storage from electrical vehicle batteries, Ref. [11] used an adaptive droop control for frequency regulation. To continuously search for optimal parameters, Ref. [12] developed an adaptive control strategy and a self-tuning algorithm for energy storage control to minimize frequency deviation and the rate of change of the frequency, while reducing the energy storage power flow. For improved robustness, fuzzy-based frequency control strategies were applied in Ref. [13] to coordinate distributed PV, energy storage and electrical vehicle batteries. A fuzzy logic controller was also applied in Ref. [14] targeting at automatic generation control (AGC) for multi-area systems.

This paper studied using energy storage to improve frequency response of power grids with high PV penetration. Two U.S. interconnection grids were studied: the EI and the ERCOT systems. High-energy-density energy storage (HEES) systems and high-power-density energy storage (HPES) systems were distinguished in this study. Two control strategies: frequency droop and step response of energy storage systems for frequency response were evaluated. Additionally, sensitivity of frequency response to key parameters of the energy storage systems, including the converter current limit, the storage capacity limit, and the discharge time, were investigated using high PV dynamic models of the U.S. EI and ERCOT systems.

## II. EI System Frequency Response Enhancement Using Energy Storage

The geographic coverage of the two interconnection grids in the U.S. is shown in Fig. 1. The EI high-PV models with various rates of PV penetration have been developed and used in various studies performed by the study team as documented in [15, 16]. PSS/e® is the simulation tool of this study. The penetration rates of PV and wind in the EI and ERCOT models used in this study are provided in Table 1. The validity of the base dynamic models without renewables have been tested by synchrophasor data of GridEye. Under different PV penetration rates, the EI system frequency responses after a 4.5 GW power loss contingency, as defined by the Resource Contingency Criteria (RCC), are shown in Fig. 2. It shows that the frequency response capability decreases as PV increases.

Fig. 3 shows the energy storage control system structure diagram. Basic information of the EI system model and key control parameters are shown in Table 2. In this study, the energy storage devices are located near PV power plants and connected to the grid in an aggregated manner. As this scoping study focuses on future power grids, these default parameters are determined empirically. For example, the maximum power of energy storage devices is close to or slightly smaller than the RCC power value, in order to fully or partially compensate the generation loss in the largest contingency. Sensitivity studies on these parameters are included in this paper.

---

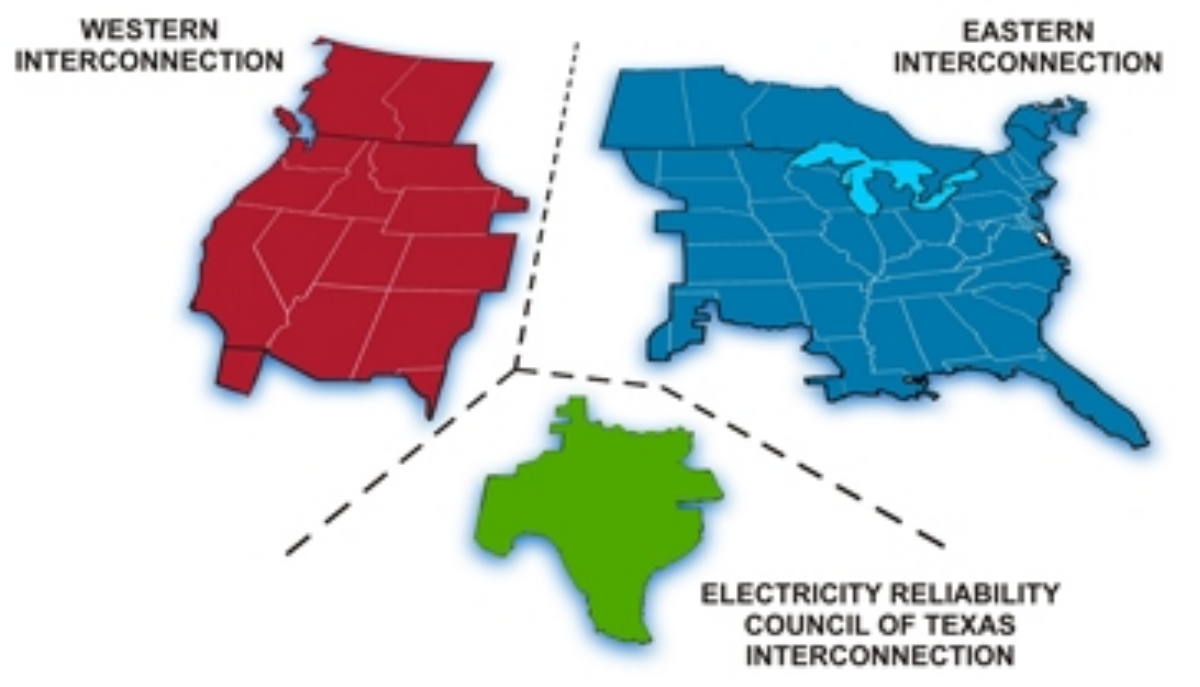

Fig. 1. The two U.S. interconnection systems (EI-blue, ERCOT-green) [17]

Table 1. PV and wind penetration in studied scenarios in the two U.S. interconnection grids

| Renewable Penetration Scenario | PV Generation Penetration Rate (%) | Wind Generation Penetration Rate (%) |
|---|---|---|
| 20% | 5 | 15% |
| 40% | 25% | 15% |
| 60% | 45% | 15% |
| 80% | 65% | 15% |

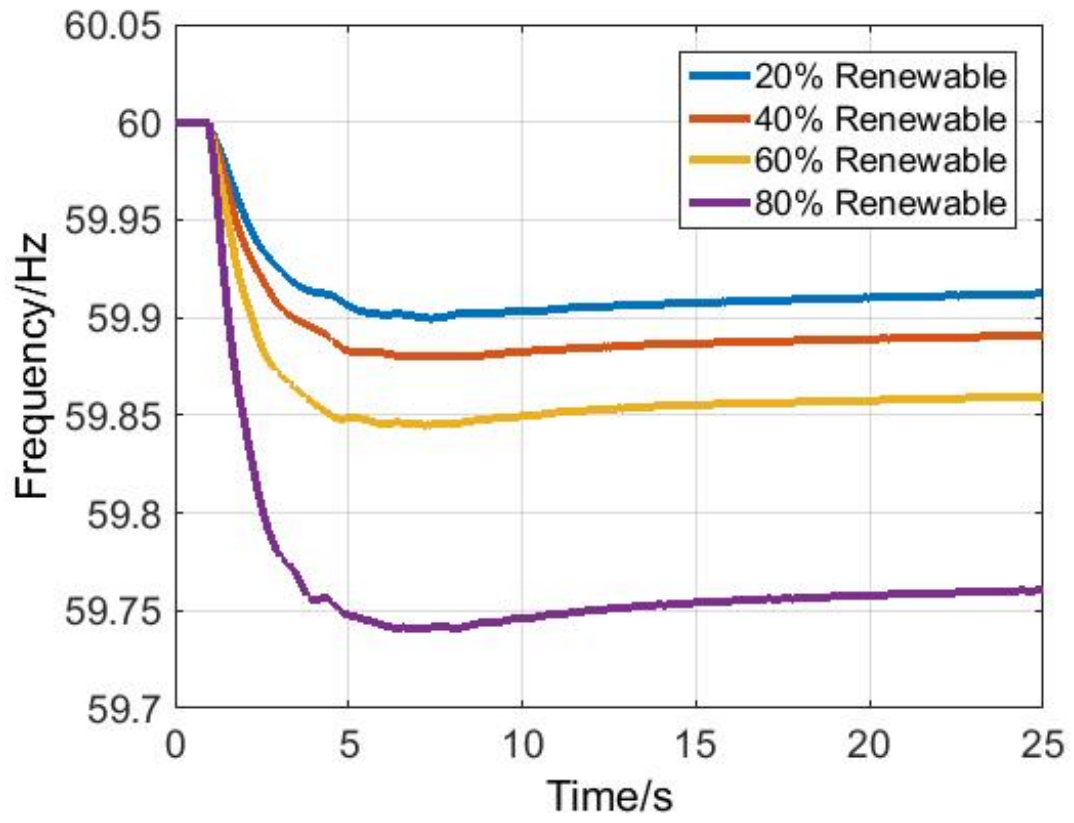

Fig. 2. Frequency response changes with the EI renewable penetration

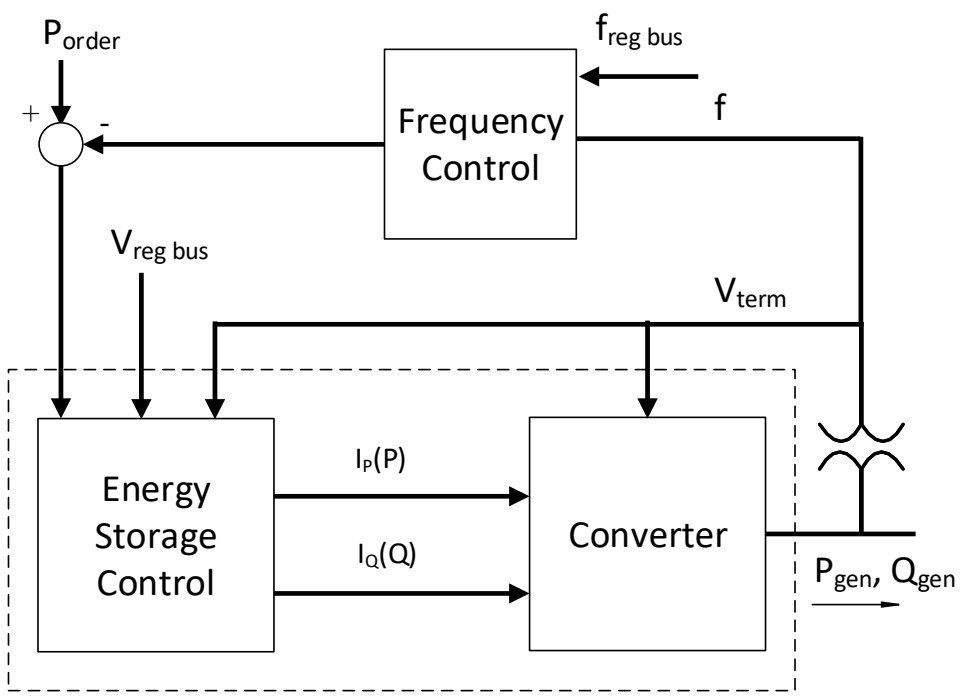

Fig. 3. Energy storage frequency control model connectivity

Table 2. The EI model information and key energy storage control parameters

| **Parameter and setting** | **Value** |
|---|---|
| Total load in EI | 560 GW |
| Bus number | ~70k |
| RCC Contingency magnitude | 4.5 GW |
| Maximum power | 3,100 MW |
| Maximum energy | 3,100 MW*10s |
| Droop frequency control ratio | 2.5% |
| Time constant of the low-pass filter $T_1$ | 0.5 s |
| Step response activation frequency | 59.85 Hz |
| Time delay in step response | 0.5 s |
| Response – contingency ratio α | 0.85 |

## *A. Droop Control and Step Response with HEES*

The discharge time of HEES systems, such as chemical batteries, varies between tens of minutes to hours. HEES systems can provide sustained power output in frequency response. The inverter current limit constrains the maximum power for frequency support. Two studied control strategies are:

*1) Droop Control:* The frequency deviation of the terminal frequency measured by phasor lock loop (PLL) is fed into a low-pass filter and a gain block is then used to control the response magnitude before adding to the initial power output point. This control principle is similar to a governor of a conventional synchronous generator. The control diagram is shown in Fig. 4.

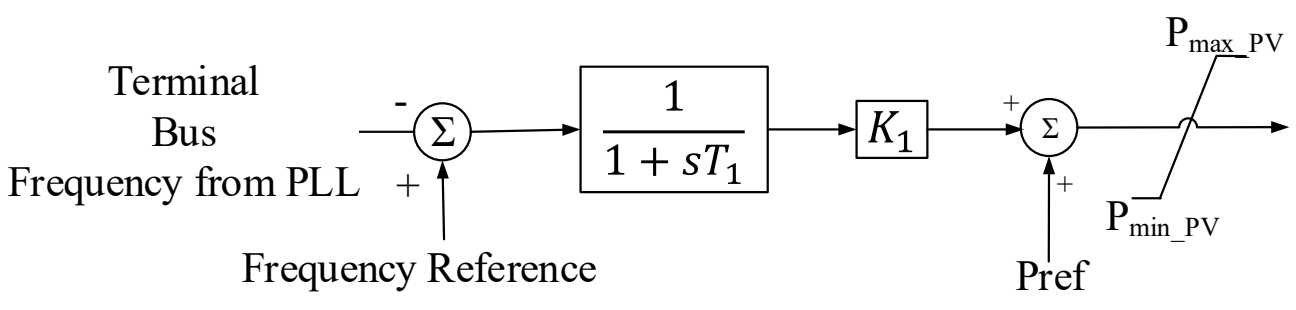

Fig. 4. Droop control of energy storage for frequency response

*2) Step Response:* This control first estimate the MW loss of the contingency using the rate of change of frequency (ROCOF) value. Then a step function is applied to control inverter real power output to support the system frequency. Step response of energy storage systems can leverage the fast response characteristics of converters. The magnitude of step response is obtained from (1).

$$P_{step} = \alpha \cdot 2 \cdot H_{sys} \cdot \frac{ROCOF}{f_N} \cdot C_{sys} \quad (1)$$

where $\alpha$ is the response coefficient ( $0 < \alpha < 1$ ), which determines the portion of generation-load imbalance to be compensated by energy storage. $H_{sys}$ is the average inertia constant of the system. $f_N$ is the rated frequency. $C_{sys}$ is the system total capacity (in MVA). A frequency threshold is adopted to activate control. A time delay is used to further confirm the contingency and calculate the contingency magnitude.

The frequency responses of the EI system for 4.5 GW generation loss by the two types of control of energy storage are plotted in Fig. 5. It shows that frequency response can be improved significantly by HEES. Compared with droop control, step response has better frequency response. This is because step response can better utilize of fast response of inverters, providing faster and stronger frequency support especially at the initial stage of frequency decline. However, as implied by the response magnitude calculation approach in (1), step response control requires accurate system condition information, including the system average time constant, the rate of change of frequency value, and the system total MVA capacity. Large errors may result in insufficient response or frequency overshoot.

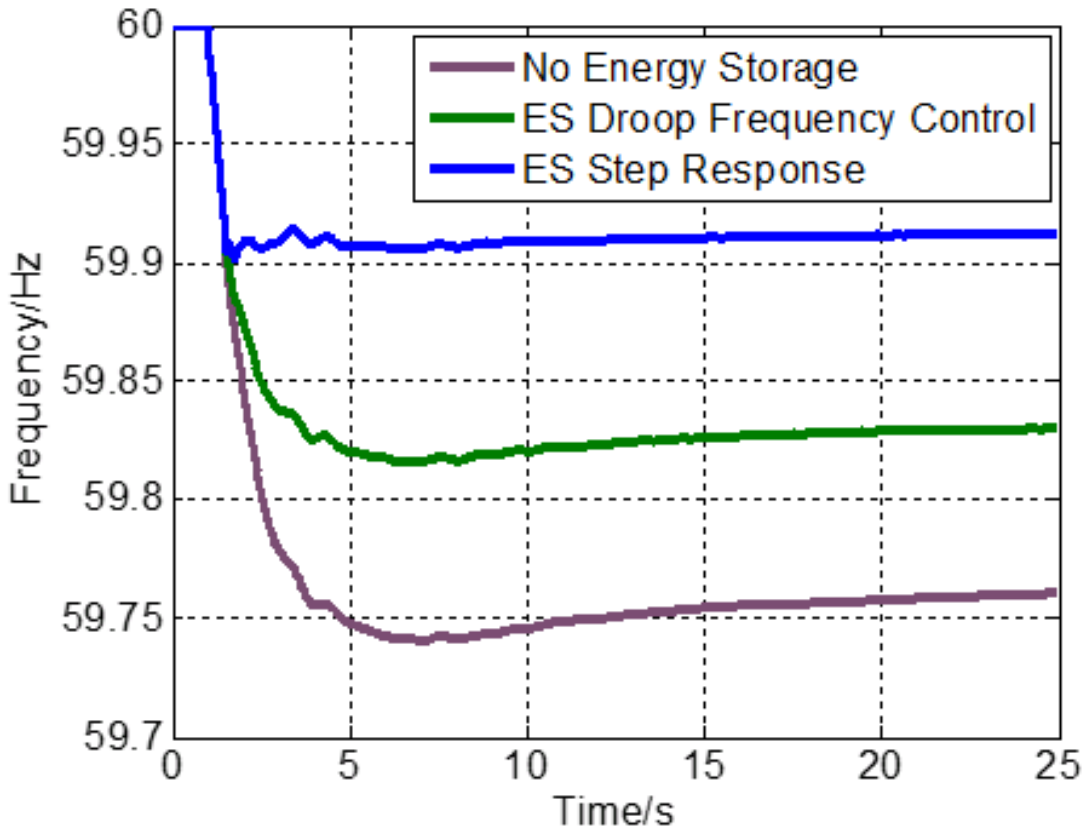

Fig. 5. Frequency responses of the EI using two types of control of HEES (4,500 MW generation loss contingency, 65% PV+15% wind)

## B. Droop Control and Step Response with HPES

HPES, such as super capacitors, can provide short-term support to frequency response due to limited energy capacities. The performances of the two controls using HPES are shown in Fig. 6. It can be noticed that a second frequency nadir will occur after exhausting stored energy in HPES. This second frequency nadir is only slightly higher than the case without HPES. This slight increase in nadir is due to the delayed response of conventional governors and turbines, which leads to stronger governor response at the time of the second frequency dip. The major benefit of HPES is delay the time of frequency reaching critical thresholds.

As a sensitivity study, Fig. 7 and Fig. 8 shows the change of the nadir frequency value and the time of nadir as the capacity of HPES changes using droop control. It can be seen that larger energy capacities of energy storage result in a higher nadir value and delayed time of nadir. It is also noted that this nadir increase is insignificant and it starts to saturate as the energy capacity further increases. In addition, when the stored energy amount is very small and only supports the system frequency for less than one second, the nadir and the nadir time are insensitive to the amount of stored energy.

The controller of HPES can adjust the discharge duration of a certain amount of stored energy by choosing different response magnitudes. Fig. 9 and Fig. 10 show the change of two frequency response metrics when the HPES varies its discharge duration (i.e. changing the response ratio α) in step response. From Fig. 9, it can be noticed that he frequency nadir will reach a peak point when the HPES is discharged for around 35s in step response. This point corresponds to the scenario in which the first and the second frequency nadir have the same frequency value. In addition, from Fig. 10, it can be seen that 35 second discharge duration also corresponds to the transition of the frequency nadir time for the first nadir to the second nadir.

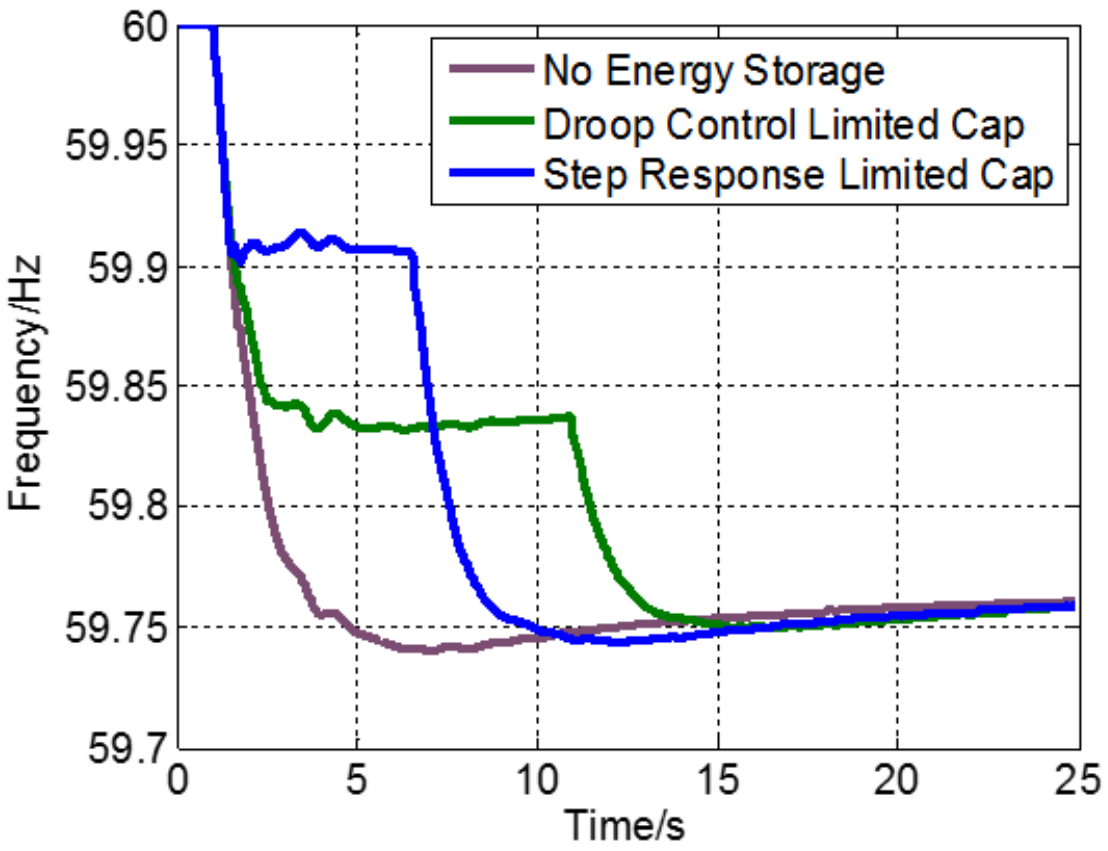

Fig. 6. Frequency responses of the EI using two types of control of HPES (4,500 MW generation loss contingency, 65% PV+15% wind)

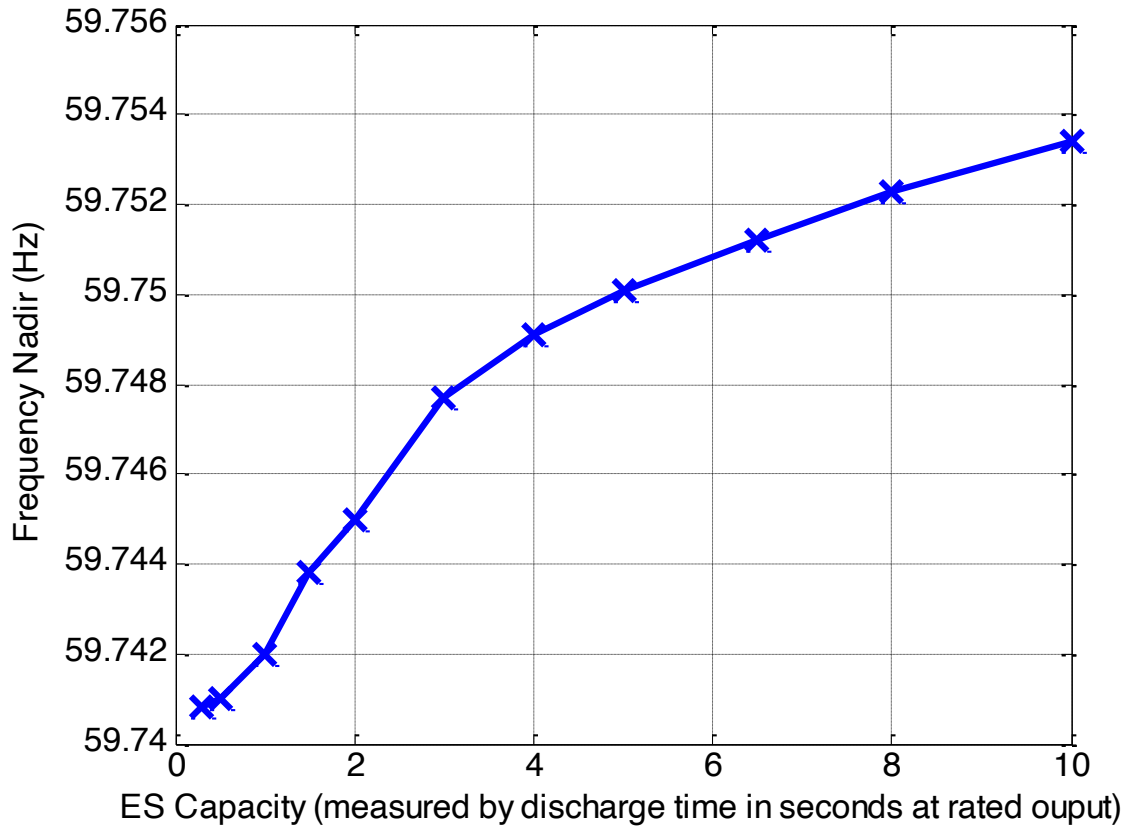

Fig. 7. Change of the frequency nadir value with the energy capacity of HPES (using droop control, the EI, 4,500 MW generation loss contingency, 65% PV+15% wind)

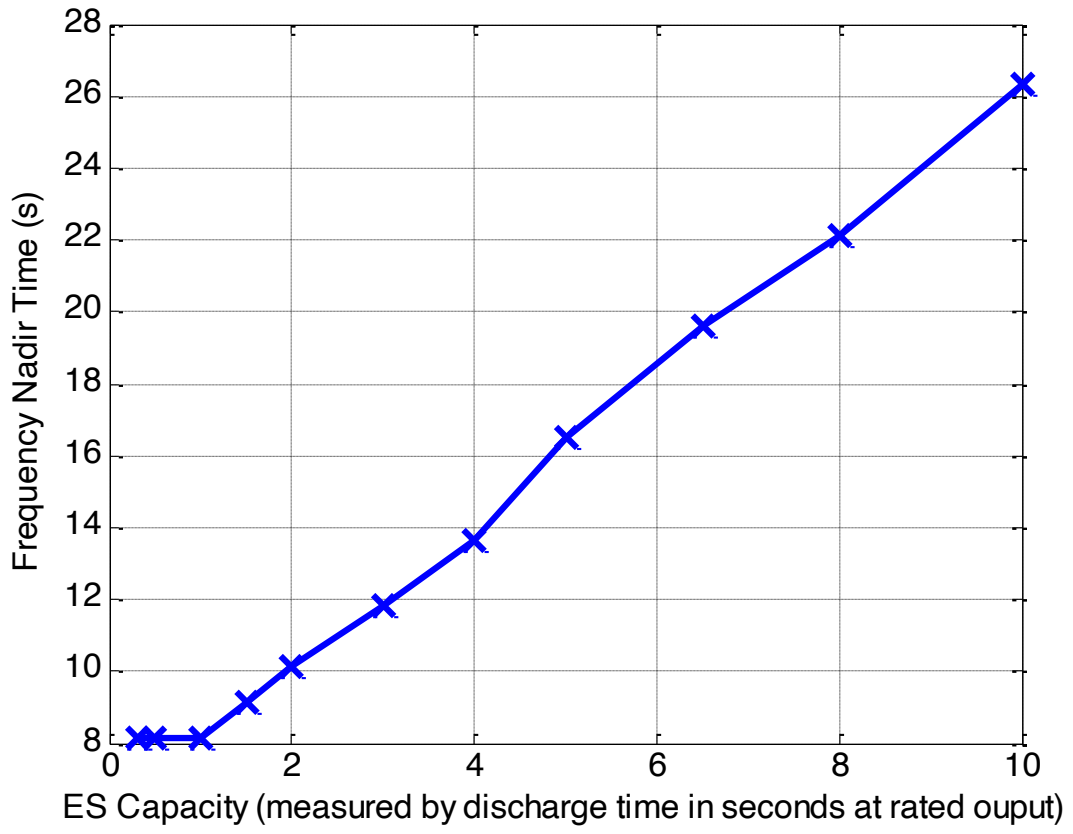

Fig. 8. Change of the time of frequency nadir with the energy capacity of HPES (using droop control, the EI, 4,500 MW generation loss contingency, 65% PV+15% wind)

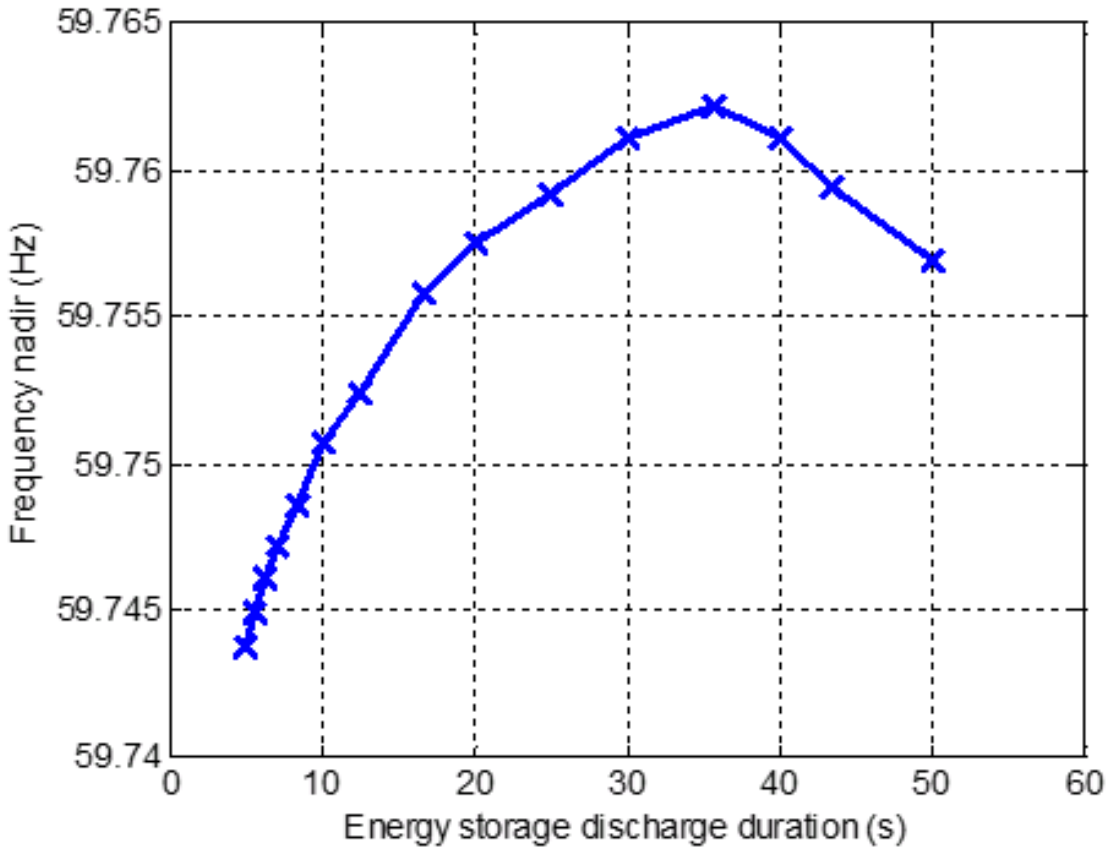

Fig. 9. Change of the frequency nadir value with the discharge duration of HPES (using step response, the EI, 4,500 MW generation loss contingency, 65% PV+15% wind)

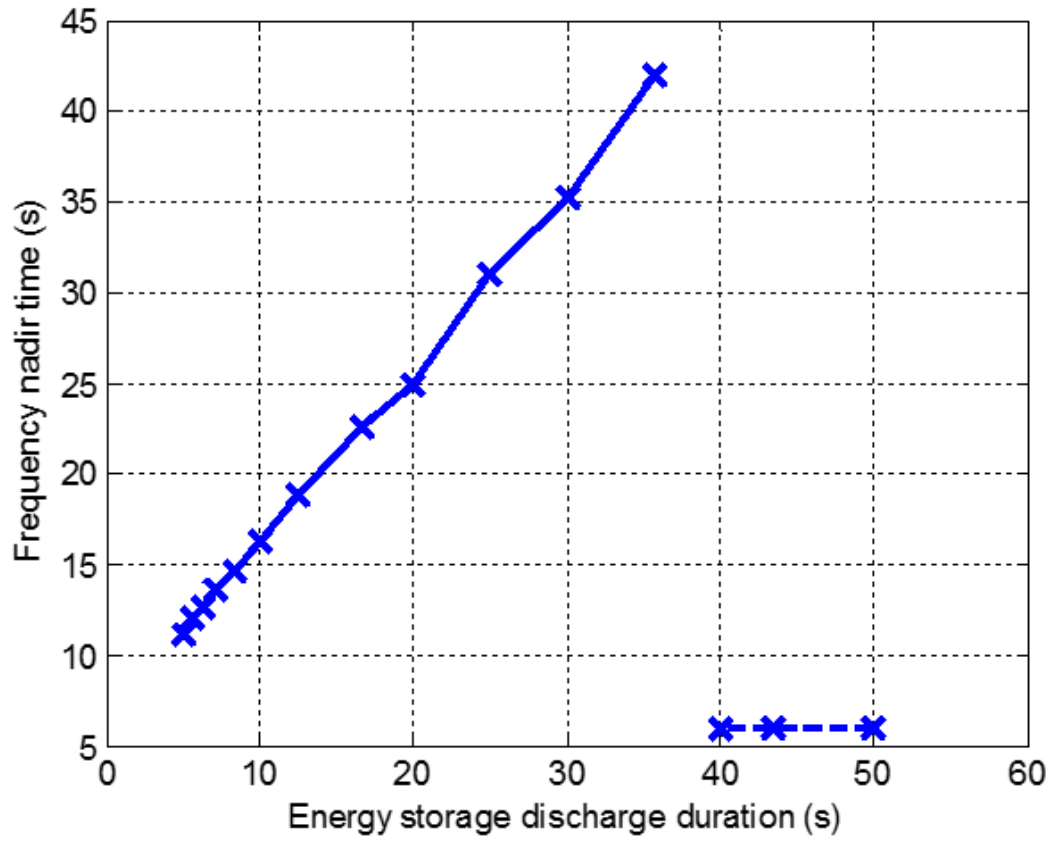

Fig. 10. Change of the frequency nadir time with the discharge duration of HPES (using step response, the EI, 4,500 MW generation loss contingency, 65% PV+15% wind)

## III. ERCOT Frequency Response Enhancement Using Energy Storage

The ERCOT system frequency responses under various PV penetration levels are plotted in Fig. 11. It shows that the system frequency response capability will not be able to prevent under-frequency-load-shedding for the 60% and 80% renewable penetration scenarios. ERCOT system model information and key parameters for energy storage control are given in Table 3.

Table 3. ERCOT model information and key energy storage control parameters

| Parameter and setting | Value |
|---|---|
| Total load in ERCOT | 75 GW |
| RCC contingency magnitude | 2,750 MW |
| Maximum power | 2,630 MW |
| Maximum energy | 2,630 MW*10 s |
| ES droop frequency control ratio | 5% |
| Time constant of the LP filter $T_1$ | 0.5 s |
| Step response activation frequency | 59.55 Hz |
| Time delay in step response | 0.5 s |
| Response – contingency ratio α | 0.85 |

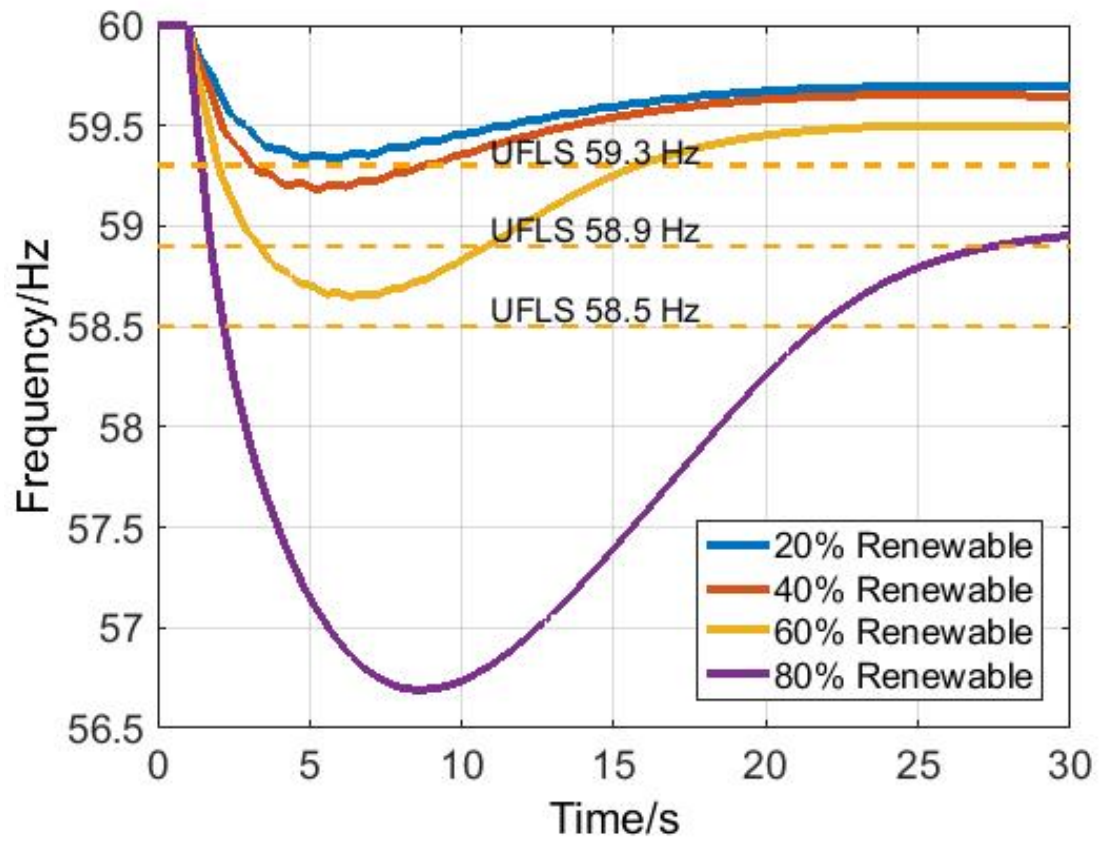

Fig. 11. Frequency response under various PV penetration scenarios in the ERCOT (2,750 MW generation loss contingency, under-frequency-load-shedding disabled, no load response considered)

### A. Droop Control and Step Response with High-Energy-Density Energy Storage (HEES)

Similar to the EI, both control approaches are implemented in the ERCOT system. Fig. 12 shows the ERCOT frequency profiles using the two control approaches in HEES. The frequency nadirs of the two controls are close: around 59.5 Hz, successfully preventing the ERCOT frequency from crossing UFLS thresholds. In addition, the step response control has an obvious frequency recovery process, due to the combined effects of the flat output of HEES and the dynamics of conventional governors. In contrast, the frequency profile is relatively flat for the droop control due to the adjustment of the output of droop control based on real-time frequency.

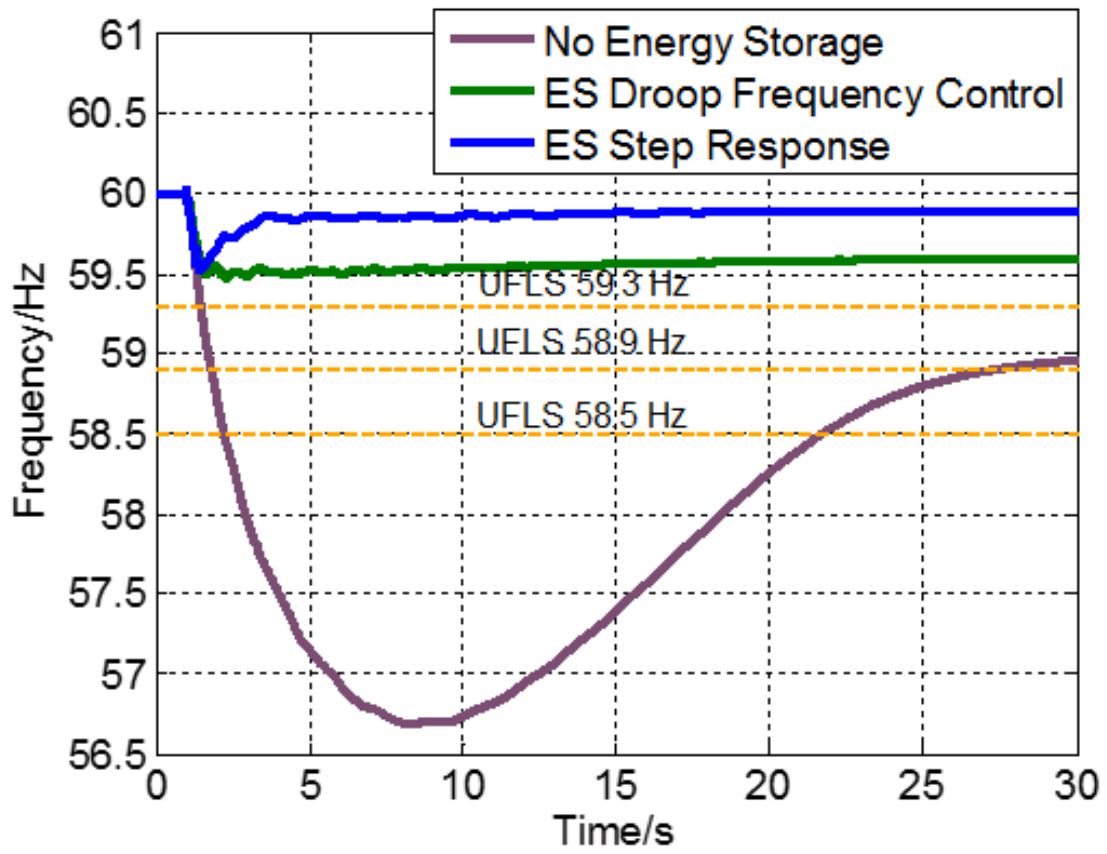

Fig. 12. The ERCOT frequency using two types of control of HEES (2,750 MW generation loss contingency, 65% PV+15% wind)

### B. Droop Control and Step Response with HPES

Energy-constrained HPES is applied to the ERCOT system. Fig. 13 shows its performance using two types of control. It can be seen that the withdrawal of energy storage results in a second frequency nadir. Step response control of HPES consumes more energy at the early stage to hold the frequency at a higher level, leading to earlier response withdrawal. Further, this earlier withdrawal of frequency support results in a lower nadir compared with that of droop frequency control.

Similar to the EI, the frequency nadir value and the nadir time also change with the energy capacity of HPES in the EROCT system, as the trends plotted in Fig. 14 and Fig. 15. It shows that if the energy of high-power-density energy storage is larger, ERCOT will have a better frequency nadir. This is because governor response becomes stronger during the second frequency dip due the time constants of governors and turbines. Because of the same reason, the ERCOT nadir can be improved slightly more compared with the EI (Fig. 7). In addition, similar to the EI, the nadir time of the ERCOT is linear to the energy storage amount.

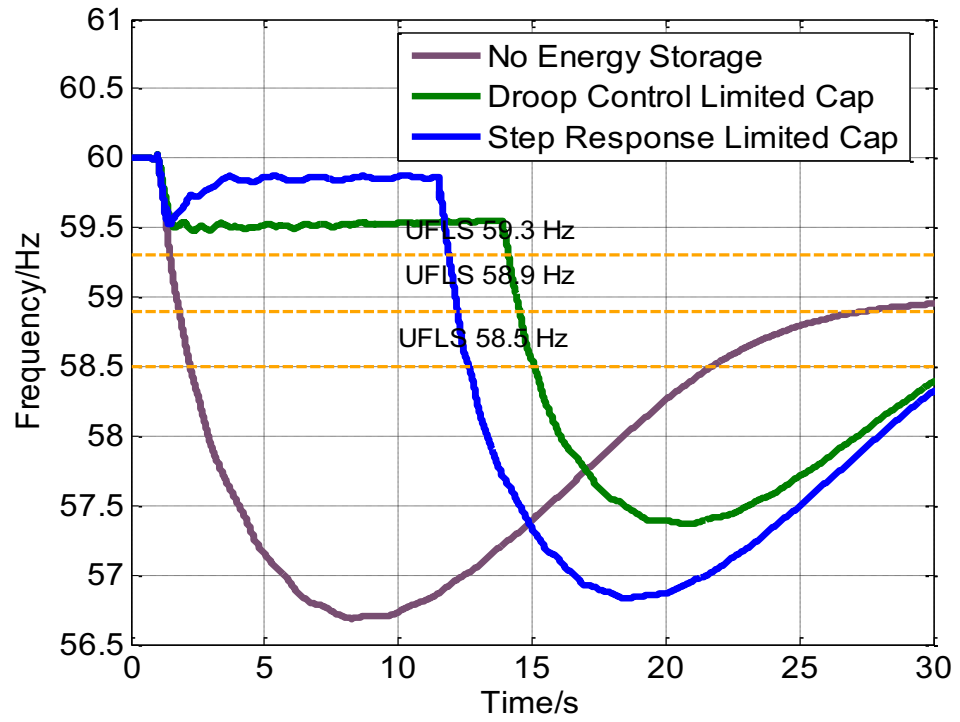

Fig. 13. Performance of two types of control of HPES in the ERCOT (2,750 MW generation loss contingency, under-frequency-load-shedding disabled, no load response considered)

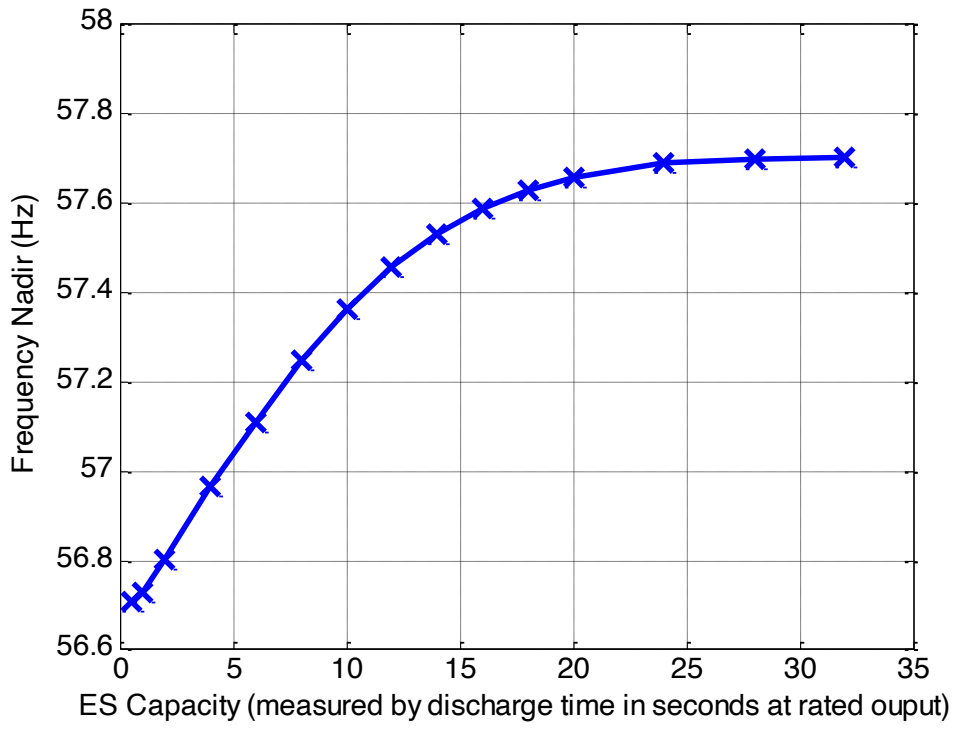

Fig. 14. Change of the nadir frequency value with the capacity of HPES in the ERCOT (droop frequency control, 2,750 MW generation loss contingency, under-frequency-load-shedding disabled, no load response considered)

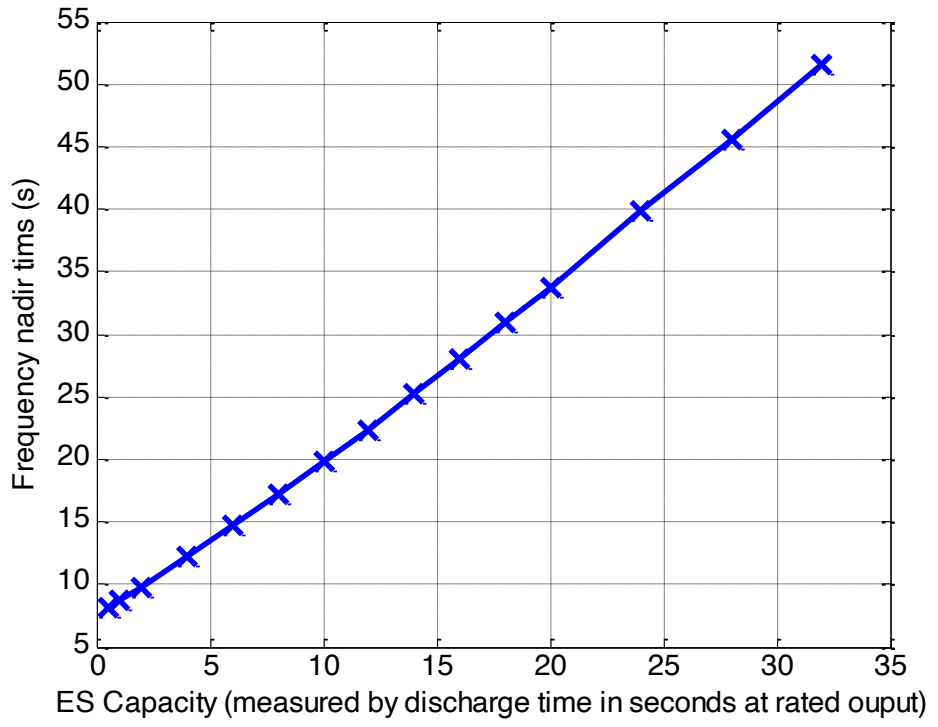

Fig. 15. Change of the nadir time with the capacity of HPES in the ERCOT (droop frequency control, 2,750 MW generation loss contingency, under-frequency-load-shedding disabled, no load response considered)

Changes of the nadir frequency value and the nadir time with the discharge duration using step response control of HPES in the ERCOT are shown in Fig. 16 and Fig. 17, respectively. The overall profile patterns are similar to those of the EI. However, the effect of different discharge duration on the nadir is more significant in the ERCOT than that in the EI. This indicates that the ERCOT has more potential to carefully tune the discharge profile of energy-constrained storage to achieve the best frequency regulation performance.

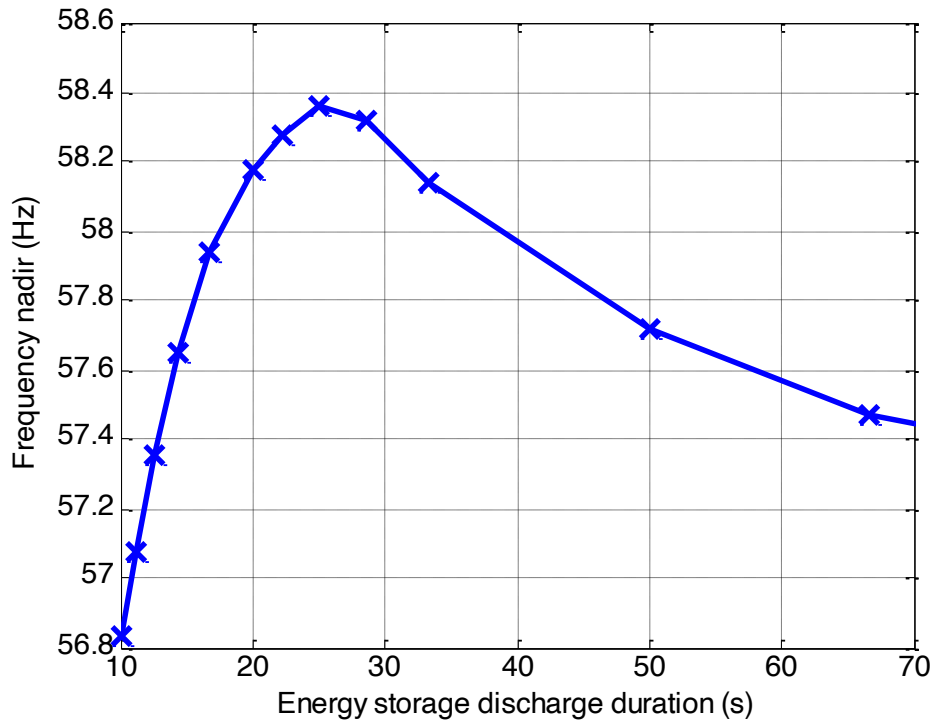

Fig. 16. Change of the nadir frequency value with the discharge duration of HPES in the ERCOT (step response control, 2,750 MW generation loss contingency, under-frequency-load-shedding disabled, no load response considered)

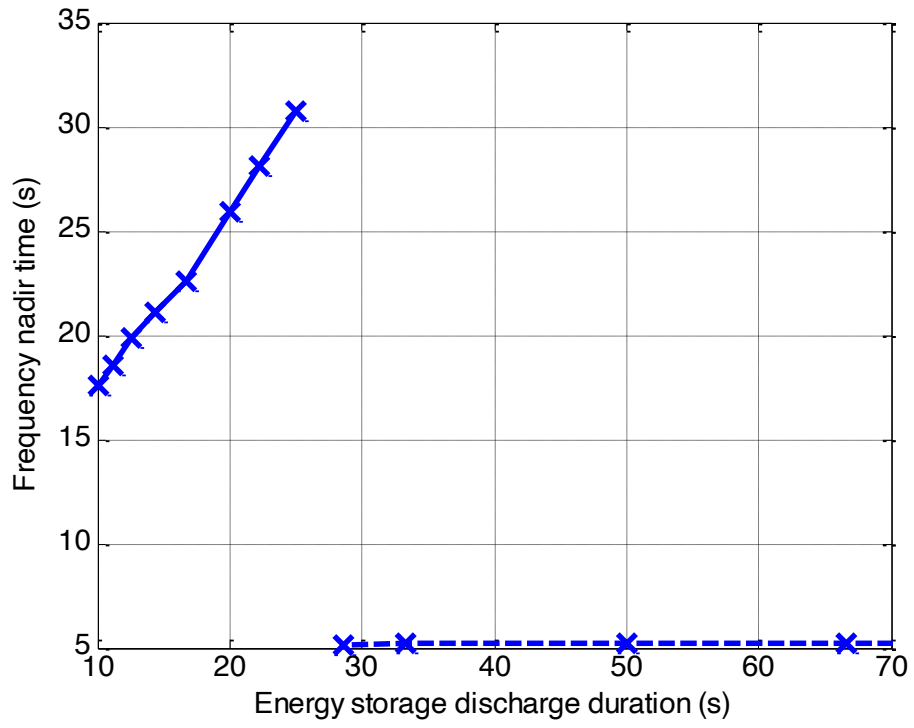

Fig. 17. Change of the nadir time with the discharge duration of HPES in the ERCOT (step response control, 2,750 MW generation loss contingency, under-frequency-load-shedding disabled, no load response considered)

## IV. Conclusions

This paper studied the performance of energy storage on interconnection-level frequency response under high PV penetration in the two U.S. interconnection systems. Results show that energy storage is effective for frequency regulation in high PV penetration. It is also found that the discharge duration and profile of energy-constrained high-power-density energy storage can be selected for optimal control performance.

## References

1. You, S., et al., *Co-optimizing generation and transmission expansion with wind power in large-scale power grids—Implementation in the US Eastern Interconnection.* Electric Power Systems Research, 2016. **133**: p. 209-218.

2. Hadley, S., et al., *Electric grid expansion planning with high levels of variable generation.* ORNL/TM-2015/515, Oak Ridge National Laboratory, 2015.
3. You, S., *Electromechanical Dynamics of High Photovoltaic Power Grids.* 2017.
4. Guo, J., et al. *An ensemble solar power output forecasting model through statistical learning of historical weather dataset*. in *2016 IEEE Power and Energy Society General Meeting (PESGM)*. 2016. IEEE.
5. Sun, K., et al., *A Review of Clean Electricity Policies—From Countries to Utilities.* Sustainability, 2020. **12**(19): p. 7946.
6. Yuan, Z., et al. *Frequency Control Capability of VSC-HVDC for Large Power Systems*. in *Power & Energy Society General Meeting, 2017 IEEE*. 2016. IEEE.
7. Datta, M., et al., *A frequency-control approach by photovoltaic generator in a PV–diesel hybrid power system.* IEEE Transactions on Energy Conversion, 2011. **26**(2): p. 559-571.
8. Mercier, P., R. Cherkaoui, and A. Oudalov, *Optimizing a battery energy storage system for frequency control application in an isolated power system.* IEEE Transactions on Power Systems, 2009. **24**(3): p. 1469-1477.
9. Delille, G., B. Francois, and G. Malarange, *Dynamic frequency control support by energy storage to reduce the impact of wind and solar generation on isolated power system's inertia.* IEEE Transactions on Sustainable Energy, 2012. **3**(4): p. 931-939.
10. Serban, I. and C. Marinescu, *Control strategy of three-phase battery energy storage systems for frequency support in microgrids and with uninterrupted supply of local loads.* IEEE Transactions on Power Electronics, 2014. **29**(9): p. 5010-5020.
11. Liu, H., et al., *Decentralized vehicle-to-grid control for primary frequency regulation considering charging demands.* IEEE Transactions on Power Systems, 2013. **28**(3): p. 3480-3489.
12. Lopes, L.A., *Self-tuning virtual synchronous machine: A control strategy for energy storage systems to support dynamic frequency control.* IEEE Transactions on Energy Conversion, 2014. **29**(4): p. 833-840.
13. Datta, M. and T. Senjyu, *Fuzzy control of distributed PV inverters/energy storage systems/electric vehicles for frequency regulation in a large power system.* IEEE Transactions on Smart Grid, 2013. **4**(1): p. 479-488.
14. ud din Mufti, M., et al., *Super-capacitor based energy storage system for improved load frequency control.* Electric Power Systems Research, 2009. **79**(1): p. 226-233.
15. You, S., et al. *Impact of High PV Penetration on U.S. Eastern Interconnection Frequency Response*. in *Power and Energy Society General Meeting (PESGM), 2017*. 2017. IEEE.
16. You, S., et al., *Impact of high PV penetration on the inter-area oscillations in the US eastern interconnection.* IEEE Access, 2017. **5**: p. 4361-4369.
17. US Department of Energy, *North American Electric Reliability Corporation Interconnections.* 2016.